\documentclass[letter,superscriptaddress,twocolumn,pra,showkeys,showpacs]{revtex4-1}

	\usepackage{amsmath}
	\usepackage{makeidx}
	\usepackage{amsfonts}
	\usepackage[ansinew]{inputenc}
	\usepackage[usenames,dvipsnames]{pstricks}
    \usepackage{graphicx}
    \usepackage{float}
    \usepackage{subfigure}
	\usepackage{pst-grad} 
	\usepackage{pst-plot} 
	\usepackage{sidecap}
	\usepackage[makeroom]{cancel}
	\usepackage{textcomp}

	\usepackage[colorlinks,hyperindex]{hyperref}
	
	\hypersetup
	{
		colorlinks,%
		citecolor=black,%
		linkcolor=black,%
		urlcolor=black,%
	}

	\newcommand{\ket}[1]{\left| #1 \right\rangle}
	\newcommand{\bra}[1]{\left\langle #1 \right|}
	
	\newcommand{\braket}[2]{\left\langle #1 \right. \left| #2 \right\rangle}

	\usepackage{color}

\makeindex

\begin{document}

\title{Entanglement dynamics of  an arbitrary number of moving qubits in a common environment}

\author{Sare Golkar}
\email{saregolkar@gmail.com}
\affiliation{Atomic and Molecular Group, Faculty of Physics, Yazd University,  Yazd, Iran}
\author{Mohammad K. Tavassoly}
\email{mktavassoly@yazd.ac.ir}
\affiliation{Atomic and Molecular Group, Faculty of Physics, Yazd University,  Yazd, Iran}
\author{Alireza Nourmandipour}
\email{anourmandip@sirjantech.ac.ir}
\affiliation{Department of Physics, Sirjan University of Technology, Sirjan, Iran}
\date{today}
\begin{abstract}
In this paper we provide an analytical investigation of the entanglement dynamics of moving qubits dissipating into a common and (in general) non-Markovian environment for both weak and strong coupling regimes. We first consider the case of two moving qubits in a common environment and then generalize it to an arbitrary number of moving qubits. We show that for an initially entangled state, the environment washes out the initial entanglement after a finite interval of time. We also show that the movement of the qubits can play a constructive role in protecting of the initial entanglement.
In this case, we observe a Zeno-like effect due to the velocity of the qubits.
On the other hand, by limiting the number of qubits initially in a superposition of single excitation, a stationary entanglement can be achieved between the qubits initially in the excited and ground states.
Surprisingly, we illustrate that when the velocity of all qubits are the same, the stationary state of the qubits does not depend on this velocity as well as the environmental properties. This allows us to determine the stationary distribution of the entanglement versus the total number of qubits in the system.

\end{abstract}

\pacs{03.65.Ud, 03.67.Mn, 03.65.Yz}
\keywords{$n$-qubit systems; Dissipative systems; Quantum entanglement; Atomic motion.}

\maketitle

\section{Introduction}

In quantum mechanics, entanglement is one of the strange behaviours of particles in which classical physics rules are broken \cite{Horodecki2009QuantumEntanglement}. It relies on the existence of correlation between individual quantum objects, such as atoms, ions, superconducting circuits, spins, or photons. Nowadays, scientists in laboratories around the world are able to entangle a large number of particles, a success that can be the basis of quantum computing \cite{nielsen2002quantum}, the technology that is expected to change the processing and storage of information in the future. It has been found that quantum entanglement is the source of many interesting applications such as quantum cryptography \cite{Ekert1991Cryptography}, quantum teleportation \cite{Braunstein1995}, superdense coding \cite{Mattle1996}, sensitive measurements \cite{Richter2007} and quantum telecloning \cite{Muarao1999}.

Many schemes have been proposed to generate the entangled states, for instance, one may refer to superconducting qubits using holonomic operations \cite{PhysRevApplied.11.014017}, trapped ions \cite{Turchette1998}, atomic ensembles \cite{Julsgaard2001}, photon pairs \cite{Aspect1981}, etc. It also has been put forward the idea that, entanglement may be generated between subsystems that never directly interacted by means of entanglement swapping \cite{Hu2011}. Recently, it
has been reported that the entanglement swapping can also be occurred between dissipative systems \cite{Nourmandipour2016swapping,ghasemi2017dissipative}.

In this regard, it ought to be emphasised that,  in real world, it is impossible to separate a quantum system from its surrounding environment. The effects of environment can destroy quantum correlations stored between subsystems, the phenomenon which is called decoherence. Therefore, instead of closed systems, we are dealt with an open quantum system \cite{Breuer2002}. At first glance, it seems quite logical to avoid as much as possible interactions with environment. However, despite the destructive effects of environment on the entanglement, it should be noticed that the environment can after all have a positive role. For instance, when two \cite{Nourmandipour2015,Maniscalco2008} or more number of qubits \cite{Nourmandipour2016,Memarzadeh2013} are interacting  with a global environment, the environment can provide indirect interaction among qubits which surprisingly lead to construct entanglement between them.  Altogether, quantum coherence results rather fragile against environment effects. Therefore, many attempts have been made to fight against the deterioration of the entanglement, for instance, quantum Zeno effect \cite{Nourmandipour2016J}, quantum feedback control \cite{Rafiee2016,Rafiee2017}, using weak measurement and measurement reversal \cite{kim2012protecting}, adding magnetic field \cite{Ghanbari2014}, etc. Furthermore, some recent studies have reported that quantum correlations can be protected by frequency modulation \cite{mortezapour2018protecting}, strong and resonant classical control \cite{gholipour2019quantumness}, dynamical decoupling pulse sequences \cite{Hu2010}, continuous driving fields \cite{Chaudhry2012} and dipole-dipole interaction \cite{golkar2019atomic}.

On the other hand,  it is impossible to consider the atoms to be static during the interaction with the electromagnetic fields in practical implementations. Therefore, it seems logical to consider the motion of the qubits. In this regard, some studies have been recently devoted to consider the motion of qubits interacting with the electromagnetic radiation \cite{Calaj2017,Moustos2017}. Specially, in an interesting work, it has been shown that in a dissipative regime, when the qubits have a uniform non-relativistic motion, they exhibit the property of preserving their initial entanglement longer than the case of qubits at rest \cite{Mortezapour2017}. This motivates us to study the possible preservation of entanglement by considering the motion of qubits in a broader context.

In this paper, we first consider the case of two moving qubits in a common environment and investigate the dynamics of entanglement in details for both strong and weak coupling regimes. We then generalize our model into a case in which an arbitrary number of qubits can interact with an environment. We shall obtain the stationary state of each case in details. We also shall illustrate that how the motion of  qubits affects the dynamics of entanglement. We also observe that, when the qubits have the same velocity, the stationary state of entanglement is completely independent of the motion of the qubits as well as the environmental parameters. We address this issue by analysing the effect of motion of the qubits on the entanglement dynamics of them.

The rest of  paper is organized as follows: In Sec. \ref{sec2} we attack the problem of two moving qubits in a common environment and obtain an analytical expression for the relevant concurrence. We investigate the entanglement dynamics in details. We do the same for an arbitrary number of qubits initially in a Werner state in Sec. \ref{sec3}. In Sec. \ref{sec4}, we consider the case in which an arbitrary number of qubits are initially in a superposition of one excitation of two arbitrary qubits. Finally, we draw our conclusion in Sec. \ref{sec5}.

\section{Two moving qubits in a common environment}\label{sec2}

\begin{figure}[ht]
   \centering
\includegraphics[width=0.4\textwidth]{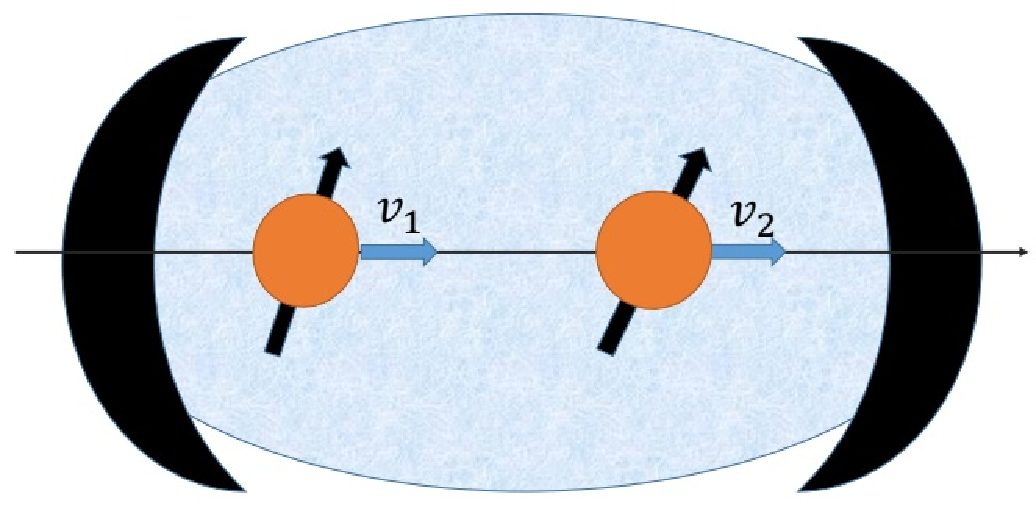}
   \caption{\label{Fig1} Pictorial representation of a setup where two qubits are moving inside a cavity. The qubits are
   two-level atoms with transition frequency $\omega_0$ travelling with constant velocity $v$.}
  \end{figure}

In this section, we consider the case in which two qubits with excited (ground) state $\ket{e}$ $(\ket{g})$ interact with a global environment (Fig. \ref{Fig1}). There is no direct interaction among these  qubits.
The qubits are taken to move along the z-axis of the cavity with constant velocity $v_i$ which can in general be different for each qubit. The environment is characterized by a spectral density of Lorentzian type. The total Hamiltonian then can be written as (we set $\hbar=1$):
 \begin{eqnarray}
 \label{eq:model}
   \hat H&=&\hat H_0+\hat{H}_{\text{int}},
  \end{eqnarray}
  in which
     \begin{eqnarray}
     \label{fig:h0}
 \hat H_0&=&\sum_{i=1}^{2}\omega_i \hat{\sigma}_+^{(i)} \hat{\sigma}_-^{(i)}+\sum_{k}\omega_{k} {\hat{a}_{k}}^{\dagger}\hat{a}_{k},
    \end{eqnarray}
    and
  \begin{eqnarray}
\hat{H}_{\text{int}}&=&\sum_{i=1}^{2}\sum_{k} \alpha_i \hat{\sigma}_+^{(i)}g_k f_k^{i}(z)\hat{a}_{k}+\text{H.c.}
\end{eqnarray}
In the above relations, $\hat{\sigma}_{\pm}^{i}$ and $\omega_i$ are the inversion operator and transition frequency of the $i$th qubit, respectively. The interaction of the $i$th qubit with the environment is measured by the dimensionless constant $\alpha_i$. The value of this parameter can be effectively manipulated by means of the Stark shifts tuning the atomic transition in and out of resonance. Furthermore, $\hat{a}_k$ and $\hat{a}_k^{\dagger}$ are the annihilation and creation operator of the $k$th mode of the environment, respectively. $g_k$ denotes the coupling constant between qubits and the $k$th mode of the environment.
The motion of the qubits is restricted in the z direction (cavity axis) \cite{PhysRevA.70.013414}. Here, we consider the general case in which the two qubits can have different velocities. In this regard, the parameter $f_k(z_{i})$ describes the shape function of the $i$th qubit motion along z-axis which is given by \cite{katsuki2013all}
\begin{equation}
f_k(z_{i})=f_k(v_{i}t)=\sin[\omega_k(\beta_it-\Gamma)] \ \ \ \  i=1,2
\label{eq:velshape}
\end{equation}
where, $\beta_i=v_i/c$ and $\Gamma=L/c$ with $L$ being the size of the cavity. The sine term in the above relation comes from the boundary conditions.

It should be noted that, the translational motion of the qubits has been treated classically $(z=vt)$. This is the situation in which the de Broglie wavelength $\lambda_B$ of the qubits is much smaller than the wavelength $\lambda_0$ of the resonant transition (i.e., $\lambda_B/\lambda_0\ll 1$) \cite{Mortezapour2017}. Which means that $\beta_i\ll 1$. Here, we assume that the two qubits can have different velocities.

It has been proven useful to introduce the collective coupling constant $\alpha_{_T}=(\alpha_1^2+\alpha_2^2)^{1/2}$ and relative (dimensionless) strengths $r_j=\alpha_j/\alpha_{_T}$. In this way, $r_1^2+r_2^2=1$ and we can only take $r_1$ as independent. Also, one can explore both the weak and strong coupling regimes by varying $\alpha_{_T}$.

For the initial state of the system, we assume that there is no excitation in the modes of the environment and the atoms in a general entangled states:
\begin{equation}
\ket{\psi_0}=\left( c_{01}\ket{e,g}+c_{02}\ket{g,e}\right)\ket{\boldsymbol{0}}_{R},
\label{eq:initialstate}
\end{equation}
in which, $\ket{\boldsymbol{0}}_{R}$ is the multi-mode vacuum state. Accordingly, the time evolution of the system is given by
\begin{equation}
\begin{aligned}
\ket{\psi(t)}&=c_1(t)e^{-i\omega_1t}\ket{e,g}\ket{\boldsymbol{0}}_{R}+c_2(t)e^{-i\omega_2t}\ket{g,e}\ket{\boldsymbol{0}}_{R} \\
 &+\sum_{k}c_{k}(t)e^{-i\omega_kt}\ket{g,g}\ket{1_k},
\end{aligned}
\label{eq:state}
\end{equation}
with $\ket{1_k}$ being the state of the environment with only one excitation in the $k$th mode.

Before discussing the general time evolution of the state Eq. (\ref{eq:state}), we notice that the reduced density operator for the two qubits in the
$\left\lbrace \ket{e,e},\ket{e,g}, \ket{g,e}, \ket{g,g}\right\rbrace $ basis is given by
\begin{equation}
 \hat{\rho}(t)= \left( \begin{array}{cccc}
0 & 0 & 0 & 0 \\
0 & \left| c_1(t)\right| ^2 & c_1(t)c_2^{*}(t) & 0 \\
0 &  c_1^{*}(t)c_2(t) &  \left| c_2(t)\right| ^2 &0 \\
0 & 0 & 0 & 1-\left| c_1(t)\right| ^2-\left| c_2(t)\right|^2 \end{array} \right).
\label{eq:densitymatrix}
\end{equation}

Upon insertion of the time-evolved state of the system (\eqref{eq:state}) in the time-dependent Schrödinger equation $\left( i\dot{\ket{\psi}}=\hat{H}\ket{\psi}\right) $, it is straightforward to observe that the equations for the probability amplitudes take the form
\begin{eqnarray}\label{coupled2}
\dot{c}_{j}(t)&=&-i \alpha_j \sum_{k} g_k f_k(z_j) c_k(t) e^{i\delta_k^{(j)}t}, \ \ \ j=1,2
\end{eqnarray}
\begin{eqnarray}\label{coupled3}
\dot{c}_{k}(t)&=&-i g_k^{*}\sum_{j=1}^{2}\alpha_j f_k(z_j)c_j(t) e^{-i\delta_k^{(j)}t},
\end{eqnarray}
where we have used $\delta_k^{(j)}=\omega_j-\omega_k$.
Without loss of generality, we assume that the two qubits have the same transition frequency, i.e., $\omega_1=\omega_2\equiv\omega_{0}$, consequently, $\delta_k^{(1)}=\delta_k^{(2)}\equiv\delta_k\equiv\omega_{0}-\omega_k$.
By integrating Eq. (\ref{coupled3}) and inserting its solution into Eq. (\ref{coupled2}), one obtains two integro-differential equations for amplitudes $c_{1}(t)$ and $c_{2}(t)$ as follow
 \begin{widetext}
\begin{equation}
\begin{aligned}
 \dot{c}_1(t)&=-\int_{0}^{t}\sum_{k}|g_k|^2e^{i\delta_k(t-t')}\left( \alpha_1^2f_k(v_1t)f_k(v_1t')c_1(t')+\alpha_1\alpha_2 f_k(v_1t)f_k(v_2t')c_2(t')\right)\text{d}t',  \\ \dot{c}_2(t)&=-\int_{0}^{t}\sum_{k}|g_k|^2e^{i\delta_k(t-t')}\left( \alpha_2^2f_k(v_2t)f_k(v_2t')c_2(t')+\alpha_1\alpha_2 f_k(v_2t)f_k(v_1t')c_1(t')\right) \text{d}t'.
\end{aligned}
\label{eq:dotus}
\end{equation}
\end{widetext}
As is observed from the above relations, the dynamics of the system depends on the velocities of the qubits. Let us consider the case in which the two qubits have the same velocity, i.e., $\beta_1=\beta_2\equiv \beta$. In this case, the equations  \eqref{eq:dotus} reduce to the following relations
\begin{subequations}
\label{eq:dotu}
\begin{eqnarray}
\dot{c}_1(t)&=&-\int_{0}^{t}F(t,t')\left( \alpha_1^2c_1(t')+\alpha_1\alpha_2 c_2(t')\right)\text{d}t',  \label{eq:u1} \\
\dot{c}_2(t)&=&-\int_{0}^{t}F(t,t')\left( \alpha_2^2c_2(t')+\alpha_1\alpha_2 c_1(t')\right) \text{d}t',  \label{eq:u2}
\end{eqnarray}
\end{subequations}
where the correlation function $F(t,t')$ reads as
\begin{eqnarray}\label{Correlation function}
F(t,t')=\sum_{k}|g_k|^2e^{i\delta_k(t-t')}f_k(vt)f_k(vt').
\end{eqnarray}
We note that according to Eqs. \eqref{eq:dotu} there exits a constant solution independently of the form of the spectral density as well as the velocity of the qubits which leads to a stationary entanglement. This long-living decoherence-free (or sub-radiant) state is obtained by setting $\dot{c}_j=0$ in Eqs. \eqref{eq:dotu}. This leads to the following state that does not decay in time
\begin{equation}
\ket{\psi_-}=r_2\ket{e,g}-r_1\ket{g,e}.
\label{eq:subradstate}
\end{equation}
As $\ket{\psi_-}$ does not evolve in time, the only relevant time evolution is that of its orthogonal, superradiant state
\begin{equation}
\ket{\psi_+}=r_1\ket{e,g}+r_2\ket{g,e}.
\label{eq:superradstate}
\end{equation}
The survival amplitude of the above state, i.e., ${\cal E}(t)\equiv \braket{\psi_+}{\psi_+(t)}$, satisfies the following relation (see Appendix \ref{App:A})
  \begin{equation}
 \dot{{\cal E}}(t)= -\alpha_T^2\int_{0}^{t}\! F(t,t'){\cal E}(t')  \, \mathrm{d}t'.
 \label{eq:surv}
  \end{equation}
It is apparent that ${\cal E}(t)$ depends on the spectral density as well as the shape function of the qubits motion. In the continuum limit for the environment, we consider the case in which the two qubits interact resonantly with a reservoir with Lorentzian spectral density $J(\omega)=W^2 \lambda/\pi[(\omega-\omega_{0})^2+\lambda^{2}]$. This is the case of two qubits interacting with a cavity field in the presence of cavity losses. Since the mirrors of the cavity are not perfectly reflective, the spectrum of the cavity field displays a Lorentzian broadening. It is possible to show that \cite{Nourmandipour2015,Maniscalco2008} for motionless qubits inside such a cavity, the correlation function $F(t,t')$ takes the form $f(t-t')=W^2e^{-\lambda t}$, with quantity $1/\lambda$ being the reservoir correlation time. For an ideal cavity (i.e. $\lambda\rightarrow 0$), $J(\omega)=W^2\delta(\omega-\omega_0)$ corresponds to a constant correlation function $f(\tau)=W^2$. In this situation, the system reduces to a two-atom Jaynes-Cummings model \cite{Tavis1968} with vacuum Rabi frequency ${\cal R}=\alpha_TW$. On the other hand, in the Markovian regime, i.e., for small correlation times (with $\lambda$ much larger than any other frequency scale), we obtain the decay rate as $\gamma=2{\cal R}^2/\lambda$. For generic parameter values, our model interpolates between these two limits.

In this situation, the correlation function \eqref{Correlation function} becomes
\begin{equation}
F(t,t')=\frac{W^2\lambda}{\pi}\int\text{d}\omega\frac{\sin[\omega(\beta t-\Gamma)]\sin[\omega(\beta t'-\Gamma)]}{(\omega-\omega_{0})^2+\lambda^{2}}e^{-i(\omega-\omega_0)(t-t')}.
\end{equation}
In the continuum limit (i.e., $\Gamma\rightarrow\infty$) \cite{park2017protection}, the analytical solution of the above relation gives rise to
\begin{equation}
F(t,t')=\frac{W^2}{2}e^{-\lambda(t-t')}\cosh[\beta\bar{\lambda}(t-t')]
\end{equation}
in which $\bar{\lambda}\equiv\lambda+i\omega_0$. As is seen, $F(t,t')=G(t-t')$. Therefore, it is quite reasonable to take the Laplace transform of both sides of Eq. \eqref{eq:surv} and transform the integro-differential equation into algebraic one which can be easily solved. Then we perform
the inverse Laplace transformation and use Bromwich integral formula \cite{Mortezapour2017Open} to find the survival
amplitude. After some straightforward but long manipulations it reads as
\begin{equation}
\label{eq:suramplit}
\begin{aligned}
{\cal E}(t)&=\frac{(q_1+y_+)(q_1+y_-)}{(q_1-q_2)(q_1-q_3)}e^{q_1\lambda t} \\
&+\frac{(q_2+y_+)(q_2+y_-)}{(q_2-q_1)(q_2-q_3)}e^{q_2\lambda t} \\
&+\frac{(q_3+y_+)(q_3+y_-)}{(q_3-q_1)(q_3-q_2)}e^{q_3\lambda t}
\end{aligned}
\end{equation}
in which the quantities $q_i \ \ (i=1,2,3)$ are the solutions of the cubic equation
\begin{equation}
\label{eq:cubic}
q^3+2q^2+(y_+y_-+\frac{R^2}{2})q+\frac{R^2}{2}=0
\end{equation}
with $y_{\pm}=1\pm\beta(1+i\omega_0/\lambda)$ and $R={\cal R}/\lambda$. Since the general cubic equation \eqref{eq:cubic} can be solved analytically, it is always possible to obtain the analytical expressions of $q_i$ and consequently find the analytical expression for ${\cal E}(t)$. However, these expressions are too long to be reported here. Once the analytical expression for ${\cal E}(t)$ is obtained, it is possible to obtain the analytical solutions for amplitudes $c_i(t)$. This can be done by introducing $\beta_{\pm}\equiv\braket{\psi_{\pm}}{\psi(0)}$ and therefore  finding  $c_j(t)$ as follow
\begin{subequations}
\label{eq:u}
\begin{eqnarray}
c_{1}(t)&=r_2\beta_-+r_1{\cal E}(t)\beta_+,  \label{eq:u1s} \\
c_{2}(t)&=-r_1\beta_-+r_2{\cal E}(t)\beta_+ .  \label{eq:u2s}
\end{eqnarray}
\end{subequations}

In what follows, we use the concurrence \cite{Wootters1998} to quantify the amount of entanglement which is defined as
\begin{equation}
{\cal C}(t):=\max\left\lbrace 0, \sqrt{\ell_1}- \sqrt{\ell_2}- \sqrt{\ell_3}- \sqrt{\ell_4}\right\rbrace,
\label{eq:con}
\end{equation}
where $\ell_j$, $j=1,2,3,4$, are the eigenvalues (in decreasing order) of the Hermitian matrix
$\hat{R}=\hat{\rho}\hat{\rho}_s$, in which $\hat{\rho}$ is the density matrix of the system and $\hat{\rho}_s=\hat{\sigma}_y\otimes\hat{\sigma}_y\hat{\rho}^{*}\hat{\sigma}_y\otimes\hat{\sigma}_y$ where $\hat{\rho}^*$ is complex conjugate of $\hat{\rho}$ in computational basis. The concurrence varies between 0 (when the qubits are separable) and 1 (when they are maximally entangled).
For the density matrix given by (\ref{eq:densitymatrix}), the concurrence becomes
\begin{equation}
{\cal C}(t)=2\left| c_1(t)c_2^*(t)\right|.  \label{eq:concurrence}
\end{equation}

\subsection{Stationary Entanglement}

Again, before discussing the general dynamics of the system, we begin by noticing that there exists a non-zero stationary value of ${\cal C}$ due to the entanglement of the decoherence-free state. First, we note that it is possible to show that if  $t\rightarrow \infty$, then ${\cal E}(t)\rightarrow 0$ for all values of the relevant parameters. Therefore, in the stationary state, $c_1=r_2\beta_-$ and $c_2=-r_1\beta_-$, which leads to a non-zero value of concurrence as
\begin{equation}
{\cal C}_s=2\left| r_1r_2\right| \left| \beta_-\right| ^2. \label{eq:stationaryconcurrence}
\end{equation}
To discuss better the time evolution of the concurrence as a function of the initial amount of entanglement stored in the system, we consider initial states of the form \eqref{eq:initialstate} with
\begin{subequations}
\label{eq:c0}
\begin{eqnarray}
c_{01}&=&\sqrt{\frac{1-s}{2}},  \label{eq:c01} \\
c_{02}&=&\sqrt{\frac{1+s}{2}}e^{i\varphi},  \label{eq:c02}
\end{eqnarray}
\end{subequations}
in which $s$ is the separability parameter with $-1\leq s\leq 1$ and $s=\pm 1$ ($s=0$) corresponds to a  separable (maximum entangled) initial state. Here, the separability parameter $s$ is related to the initial concurrence as $s^2=1-{\cal C}(0)^2$.

\begin{figure}[htp]
\centering
\subfigure[\label{fig2a} \ $\varphi=0. $]{\includegraphics[width=0.5\textwidth]{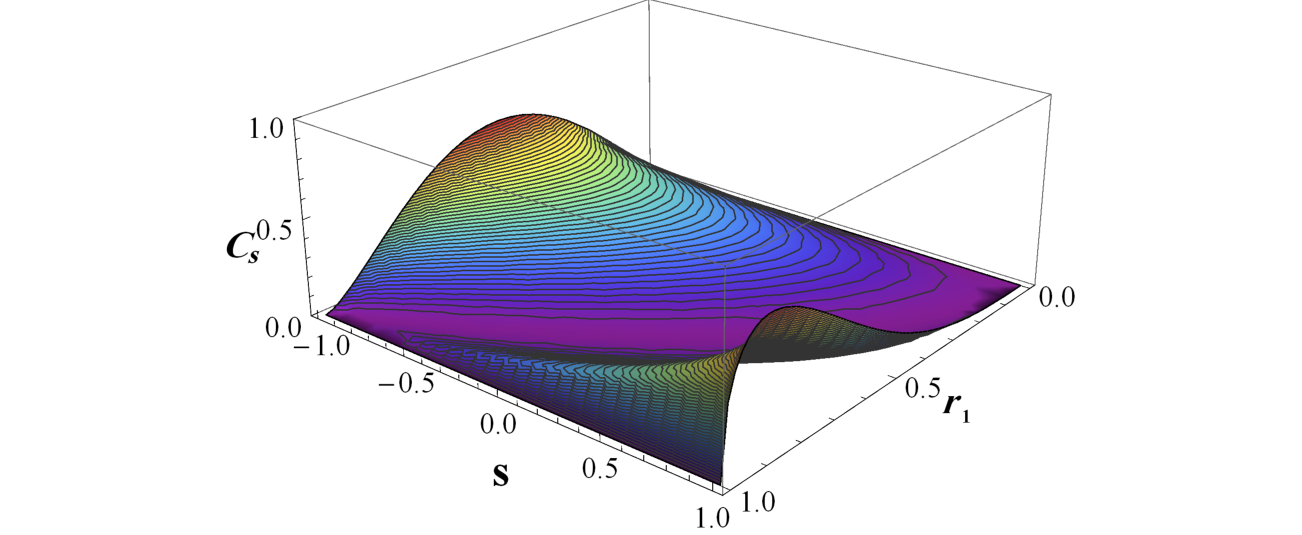}}
\hspace{0.0001\textwidth}
\subfigure[ \label{fig2b} \ $\varphi=\pi.$]{\includegraphics[width=0.5\textwidth]{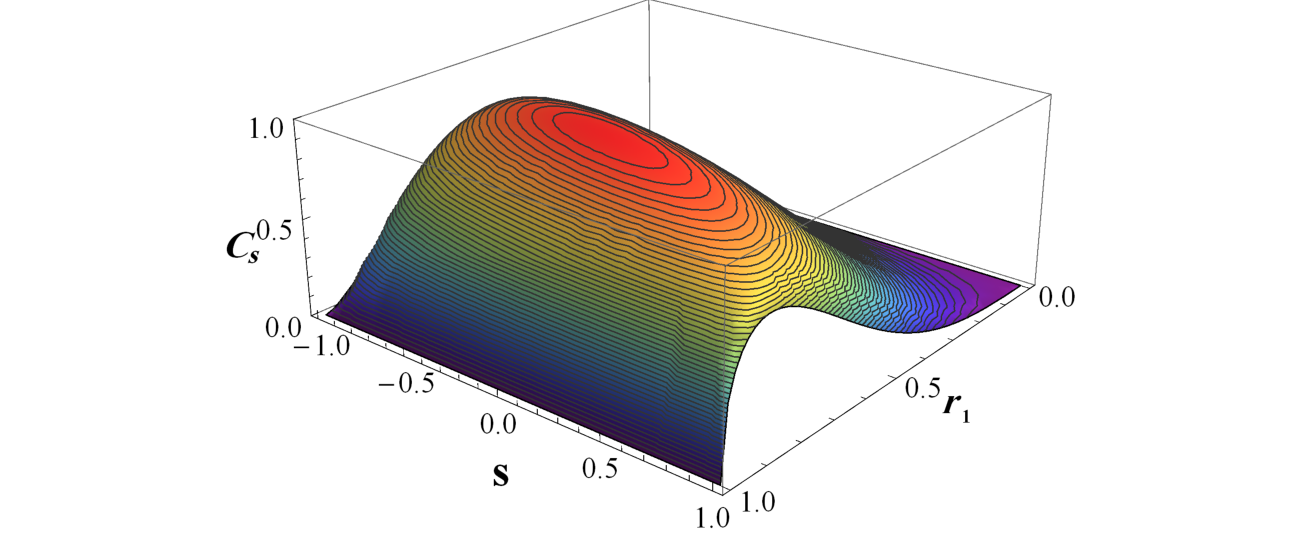}}

\caption{\label{fig2} Stationary concurrence as a function of the relative coupling constant $r_1$ and the initial separability parameter $s$ for (a) $\varphi=0$ and (b) $\varphi=\pi$.}
\end{figure}

The surprising aspect here is that, the stationary state of the two moving qubits with the same velocities (i.e., Eq. \eqref{eq:stationaryconcurrence}) is exactly the same as  the stationary state of two motionless qubits which has been reported in \cite{Nourmandipour2015,Maniscalco2008}.
However, it is useful to discuss the stationary entanglement. In Fig. \ref{fig2} we have plotted the stationary entanglement as function of the relative coupling constant $r_1$ and the initial separability parameter $s$ for two values of $\varphi$, i.e., $\varphi=0$ and $\pi$. It can clearly be seen that, for $r_1=0$ and 1, there is no stationary entanglement as these  correspond to cases in which only one atom interacts with the environment and there is no correlation between qubits due to the environment. In the case $\varphi=0$, the maximum stationary entanglement $C_{\text{s}}^{\text{max}}\approx 0.65$ is achievable for factorized initial states, i.e., this value is obtained at $r_1 = 0.5$ for $s =-1$ and at $r_1 = 0.87$
for $s = +1$. In the case $\varphi= \pi$, the maximum value of the stationary entanglement
$C_{\text{s}}^{\text{max}}=1$ is obtained at $r_1\approx 0.7$ for the maximum entanglement of the initial state $(s = 0)$, which according to Eq. \eqref{eq:subradstate}, this maximum is achieved for $\ket{\psi_0}=\ket{\psi_-}$. We point out
that the results are independent of the velocities of qubits as well as the structure of the environment.

\subsection{Dynamics of Entanglement}

Now we are in a position to discuss the entanglement dynamics of  moving qubits versus the dimensionless (scaled) time $\tau=\lambda t$. The parameter $R={\cal R}/\lambda$ allows us to consider two distinct regimes for cavity, namely good and bad cavity limits with $R\gg 1$ and $R\ll 1$, respectively.  By good cavity, we mean the case in which non-Markovian dynamics occurs. This is accompanied by an oscillatory reversible decay and the memory effect of the cavity appears. While, in the bad cavity limit, the behaviour of the atom-reservoir system is Markovian with irreversible decay in which all the history is forgotten \cite{Bellomo2007}.

\begin{figure}
\centering
\subfigure[\label{Fig3a} \ Factorized initial state, $s=1$.]{\includegraphics[width=0.23\textwidth]{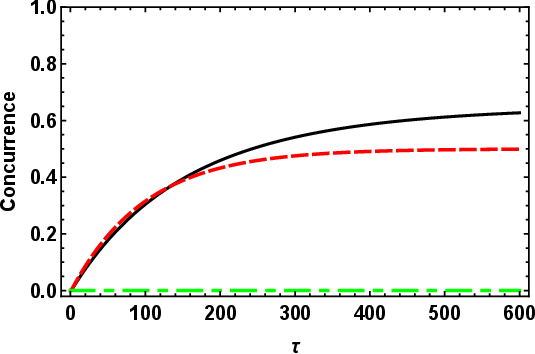}}
\hspace{0.001\textwidth}
\subfigure[\label{Fig3b} \ Entangled initial state, $s=0$.]{\includegraphics[width=0.23\textwidth]{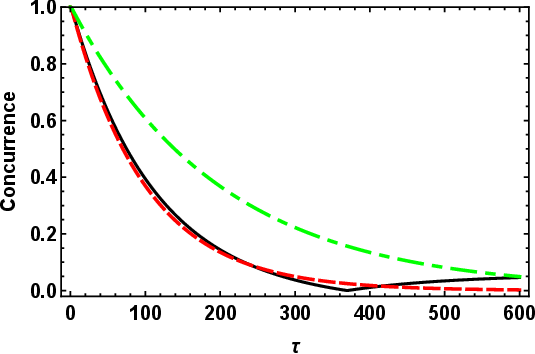}}
 \hspace{0.001\textwidth}
\subfigure[\label{Fig3c} \ Factorized initial state, $s=1$.]{\includegraphics[width=0.23\textwidth]{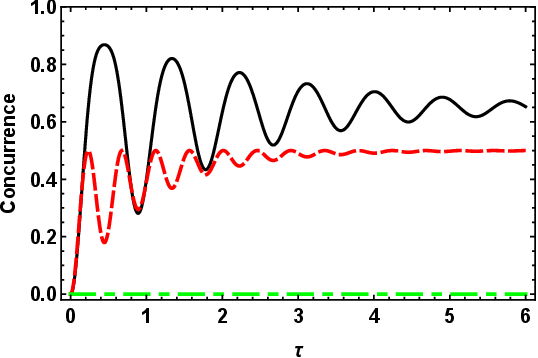}}
\hspace{0.001\textwidth}
\subfigure[\label{Fig3d} \ Entangled initial state, $s=0$.]{\includegraphics[width=0.23\textwidth]{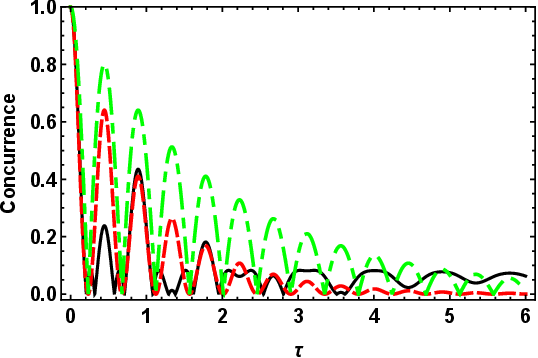}}

\caption{Concurrence of motionless qubits ($\beta=0$) as function of scaled time $\tau$ for $\varphi=0$ in the bad cavity limit, i.e. $R=0.1$ (top plots) and good cavity limit, i.e., $R=10$ (bottom plots) with $s=1$ (left plots) and $s=0$ (right plots) with the cases ($i$) maximal stationary value, $r_1=0.87$ (solid line), ($ii$) symmetric coupling, $r_1=1/\sqrt 2$ (dashed line), and ($iii$) only one coupled atom, $r_1=0$ or $r_1=1$ (dot-dashed line).} \label{Fig3}
   \end{figure}

In Fig. \ref{Fig3} we show the concurrence as a function of $\tau$ for motionless qubits (i.e., $\beta=0$) in the bad (upper row) and good (lower row) cavity limits for $\varphi=0$. In this case, we recover the results presented in \cite{Nourmandipour2015,Maniscalco2008}. However, it is worthwhile to mention that for both good and bad cavity limits and for an initially separable state $(s=1)$, the concurrence starts from zero and reaches to its stationary value. However, in the bad cavity limit, the concurrence increases monotonically up to this value, whereas, in the good cavity limit we observe entanglement oscillations and revival phenomena for every initial qubit state. This is due to the memory depth of the reservoir. Actually, the reservoir correlation time is greater than the relaxation time and non-Markovian effects become dominant. In our case, the amount of revived entanglement is huge and is comparable to previous maximum. For initially entangled state, the concurrence starts from its maximum value and decreases until it vanishes. Again, in the good cavity limit, the oscillations of the concurrence is clearly seen.

\begin{figure}
\centering
\subfigure[\label{Fig4a} \ Entangled initial state, $s=0$.]{\includegraphics[width=0.23\textwidth]{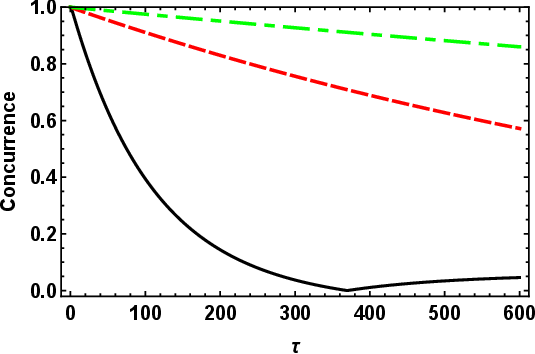}}
\hspace{0.001\textwidth}
\subfigure[\label{Fig4b} \ Factorized initial state, $s=1$.]{\includegraphics[width=0.23\textwidth]{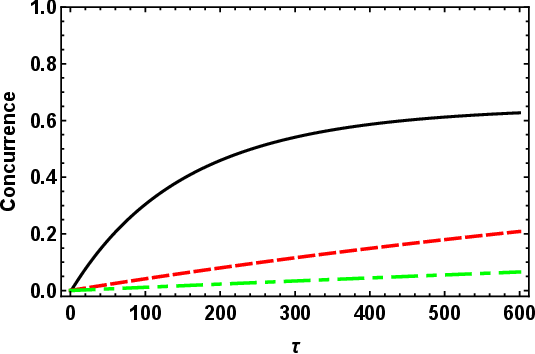}}

\caption{Concurrence as function of scaled time $\tau$ for $\varphi=0$ in the bad cavity limit, i.e. $R=0.1$ with $s=0$ (left plots) and $s=1$ (right plots). In these plots, we have set $r_1=0.87$, $\omega_0/\lambda=1.5\times 10^9$ and ($i$)  $\beta=0$ (solid line), ($ii$)  $\beta=2\times 10^{-9}$ (dashed line), and ($iii$)  $\beta=4\times 10^{-9}$ (dot-dashed line).} \label{Fig4}
   \end{figure}

In Fig. \ref{Fig4} we have plotted the dynamical behaviour of concurrence against the scaled time  in the bad cavity limit, i.e., $R=0.1$ for non-zero values of velocity. In these plots, we have set $r_1=0.87$ and $\omega_0/\lambda=1.5\times 10^9$. As is seen, for an initially entangled state, i.e., $s=0$, the movement of the qubits has a remarkably effect on the survival of the initial entanglement. As the velocity of qubits is increased, the entanglement survives in long times. On the other hand, for a factorized initial state, the qubit movement has a constructive role on the entanglement, in the sense that it makes the entanglement reaches its stationary value in longer times. It should be noted that the stationary value is independent of the  qubits velocity.

\begin{figure}
\centering
\subfigure[\label{Fig5a} \ Entangled initial state, $s=0$.]{\includegraphics[width=0.23\textwidth]{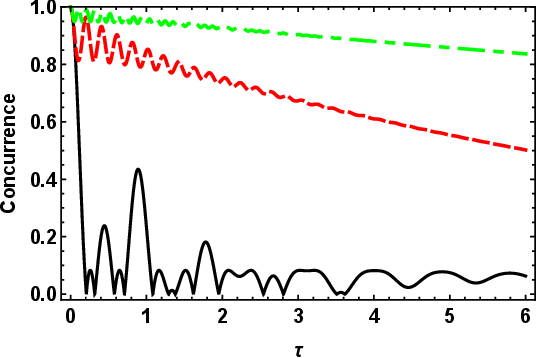}}
\hspace{0.001\textwidth}
\subfigure[\label{Fig5b} \ Factorized initial state, $s=1$.]{\includegraphics[width=0.23\textwidth]{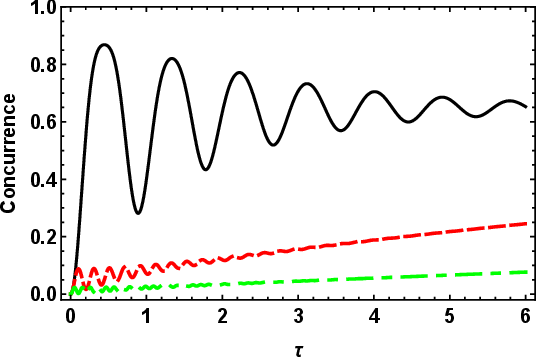}}

\caption{Concurrence as a function of scaled time $\tau$ for $\varphi=0$ in the good cavity limit, i.e. $R=10$ with $s=0$ (left plots) and $s=1$ (right plots). In these plots, we have set $r_1=0.87$, $\omega_0/\lambda=1.5\times 10^9$ and ($i$)  $\beta=0$ (solid line), ($ii$)  $\beta=2\times 10^{-9}$ (dashed line), and ($iii$)  $\beta=4\times 10^{-9}$ (dot-dashed line).} \label{Fig5}
   \end{figure}

Figure \ref{Fig5} shows the concurrence in the good cavity limit, i.e., $R=10$ for non-zero values of velocity. Again,
we have set $r_1=0.87$ and $\omega_0/\lambda=1.5\times 10^9$. As is seen, in the presence of velocity, the oscillating behaviour of entanglement washes out. Similar to the weak coupling regime, the movement of qubits has a remarkable effect on the surviving of the entanglement.  Again, the stationary value is independent of the velocity of qubits.

\section{An arbitrary number of moving qubits  INITIALLY IN A WERNER STATE}
\label{sec3}

In this section, we apply the same process for an arbitrary numbers of moving qubits which initially prepared in a Werner state. The total Hamiltonian is described as
 \begin{equation}
 \begin{aligned}
   \hat H&=\omega_0\sum_{i=1}^{n} (\hat{\sigma}_+^{(i)} \hat{\sigma}_-^{(i)})+\sum_{k}\omega_{k} {\hat{a}_{k}}^{\dagger}\hat{a}_{k}\\
   &+\alpha_{_T}\sum_{i=1}^{n}\sum_{k} \;\hat{\sigma}_+^{(i)} g_k f_k(z)\hat{a}_{k}+\text{H.c.}
   \end{aligned}
  \end{equation}
where we have assumed that the qubits have the same transition frequency $\omega_0$ and that the parameter $\alpha$ which determines the interaction of each qubit with the environment, be the same for every qubit, (i.e., $\alpha_1=\alpha_2=\cdots\alpha_n\equiv\alpha_{_T}$). We also have considered the case in which the velocity of each qubit is the same.
Suppose the initial state of the system is
\begin{equation}
\label{eq:inin}
\ket{\psi(0)}=\ket{w}\ket{\boldsymbol{0}}_{R},
\end{equation}
where  $\ket{w}:=\dfrac{1}{\sqrt{n}}\sum_{i=1}^{n}\ket{e_{i}}$ is the Werner state such that
$\ket{e_{i}}=\ket{g_1,...,e_i,...,g_n}$.
The initial state \eqref{eq:inin} evolves into state
\begin{equation}
\label{eq:statewerner}
\ket{\psi(t)}={\cal D}(t)e^{-i\omega_0t} \ket{w}\ket{\boldsymbol{0}}_{R}+\sum_{k}\Lambda_{k}(t)e^{-i\omega_k t}\ket{1_k}\ket{G}
\end{equation}
in which $\ket{G}:=\ket{g}^{\otimes n}$ and
\begin{equation}
\left| {\cal D}(t)\right| ^2\equiv\text{P}_0(t)=\left| \braket{\psi(0)}{\psi(t)}\right| ^2
\label{eq:suramp}
\end{equation}
is the survival probability  (fidelity) of the initial state.
Following the same procedure as is done for the two-qubit case, we are readily led to  integro-differential equation for the amplitude ${\cal D}(t)$:
\begin{eqnarray}\label{cc1}
\dot{{\cal D}}(t)&=&-n\alpha_T^2 \int_{0}^{t}   F(t,t'){\cal D} (t') \ \mathrm{d}t',
\end{eqnarray}
in which the correlation function $F(t,t')$ has been introduced in Eq. \eqref{Correlation function}. Therefore, the amplitude ${\cal D}(t)$ is obtained like Eq. \eqref{eq:suramplit}, with the quantities $q_i$ are now the solutions of the following cubic equation:
\begin{equation}
\label{eq:cubic2}
q^3+2q^2+(y_+y_-+\frac{nR^2}{2})q+\frac{nR^2}{2}=0.
\end{equation}

 Again it is readily observed that at sufficiently long times (i.e., $t\rightarrow\infty$), ${\cal D}(\infty)\longrightarrow 0$. Therefore, looking at (\ref{eq:statewerner}), one can realise that $\ket{\psi(\infty)}\propto \ket{G}$, which implies that when all qubits are initially in a superposition of single excited states with the same probability, no stationery entanglement can be achieved.

Using (\ref{eq:statewerner}) the explicit form of the reduced density operator for the system of qubits can be derived by tracing over environment variables. Then, in order to analyse the pairwise entanglement between any two generic qubits, we compute partial trace of the resulted density matrix over all other qubits and obtain the following reduced density operator:
\begin{equation}
\rho_{\text{pair}}(t)=\begin{pmatrix}
 0 & 0 & 0 & 0 \\
 0 & \dfrac{\left| {\cal D}(t)\right| ^2}{n} & \dfrac{\left| {\cal D}(t)\right| ^2}{n} & 0 \\
 0 & \dfrac{\left| {\cal D}(t)\right| ^2}{n} & \dfrac{\left| {\cal D}(t)\right| ^2}{n} & 0 \\
 0 & 0 & 0 & 1-\dfrac{2\left| {\cal D}(t)\right| ^2}{n}
 \end{pmatrix},
 \label{eq:rwkl}
\end{equation}
where the relevant concurrence reads as ${\cal C}_{\text{pair}}(t)=2\left| {\cal D}(t)\right| ^2/n$.  Keeping in mind (\ref{eq:suramp}), it is readily found that ${\cal C}_{\text{pair}}(t)=2\text{P}_0(t)/n$, which implies that the pairwise concurrence directly depends on the survival probability of the initial state.

\begin{figure}[h!]
\centering
\subfigure[\label{Fig6a} \ Bad cavity limit, $R=0.1$.]{\includegraphics[width=0.23\textwidth]{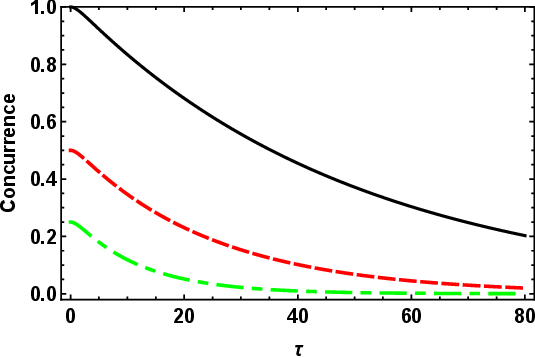}}
\hspace{0.001\textwidth}
\subfigure[\label{Fig6b} \ Good cavity limit, $R=10$.]{\includegraphics[width=0.23\textwidth]{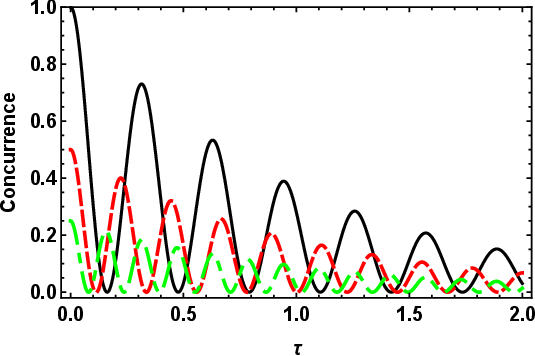}}

\caption{(Color online) Pairwise concurrence ${\cal C}_{\text{pair}}$ as function of $\tau$ when the initial state of the system is a Werner state, in the bad cavity limit, i.e. $R=0.1$ (left plots) and good cavity limit, $R=10$ (right plots) with $n=2$ (solid black line), $n=4$ (dashed red line), and $n=8$ (dot-dashed green line).} \label{Fig6}
   \end{figure}

 In this line, two distinct coupling regimes, i.e., weak and strong can be distinguished. The quantity ${\cal C}_{\text{pair}}(\tau)$ is shown in Fig. \ref{Fig6} in both regimes when the qubits are at rest. In weak coupling regime, the behaviour of concurrence is essentially a Markovian exponential decay. The concurrence disappears faster when the system size becomes larger. The strong coupling represents the revival and oscillation of entanglement. This revival phenomenon is due to the long memory of the reservoir.  In this case, the reservoir correlation time is greater than the relaxation time and non-Markovian effects become dominant. No stationary entanglement is seen for both coupling regimes. It means that, at sufficiently long times, we are left with an ensemble of non-correlated qubits.

\begin{figure}[h!]
\centering
\subfigure[\label{Fig7a} \ Bad cavity limit, $R=0.1$.]{\includegraphics[width=0.23\textwidth]{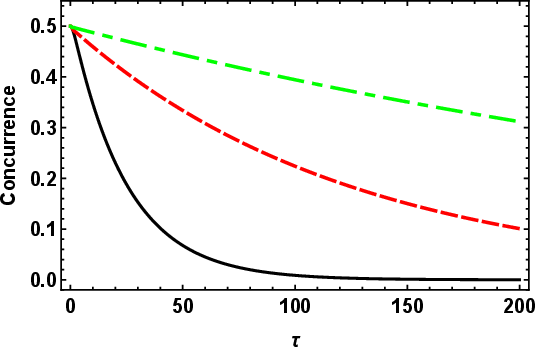}}
\hspace{0.001\textwidth}
\subfigure[\label{Fig7b} \ Good cavity limit, $R=10$.]{\includegraphics[width=0.23\textwidth]{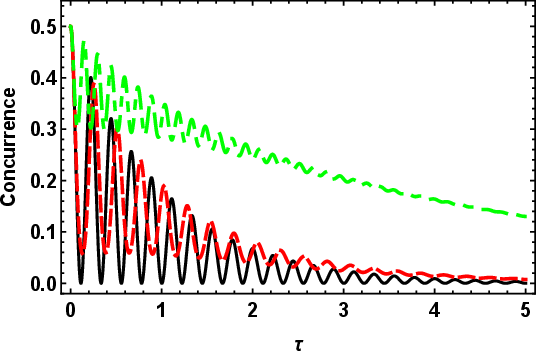}}

\caption{(Color online) Pairwise concurrence ${\cal C}_{\text{pair}}$ as function of $\tau$ when the initial state of the system is a Werner state, in the bad cavity limit, i.e. $R=0.1$ (left plots) and good cavity limit, $R=10$ (right plots). In these plots, we have set $n=4$, $\omega_0/\lambda=1.5\times 10^9$ and ($i$)  $\beta=0$ (solid line), ($ii$)  $\beta=2\times 10^{-9}$ (dashed line), and ($iii$)  $\beta=4\times 10^{-9}$ (dot-dashed line). } \label{Fig7}
   \end{figure}

In Fig. \ref{Fig7}, we have plotted the parameter ${\cal C}_{\text{pair}}$ when the initial state of the system is a Werner state with system size $n=4$ in the presence of movement of qubits. The results seem to prove that the movement of the qubits has a remarkable effect on the survival of the initial entanglement. As the velocity increases, the entanglement survives at longer times. Again, in the good cavity limit, the oscillating behaviour of the pairwise concurrence is washing out. Based on these results, one can think of an entanglement protection. Actually, the movement of  qubits plays an entanglement protection role. In the next section, we examine the suggested process for the case in which a finite number of qubits (two or one) are initially excited.

\section{System of  $n$-Qubits Initially in a superposition of one excitation of two arbitrary qubits }\label{sec4}
In this section, we  address the case in which the initial state of the qubits is a superposition of one excitation of two arbitrary qubits, namely the $j$th and $l$th qubits. Again, we assume that there is no excitation in the cavity before the occurrence of interaction.  Therefore, the initial state of the whole system+environment can be written as
\begin{equation}
\ket{\psi(0)}=(c_{01}\ket{e_j}+c_{02}\ket{e_l})\ket{\boldsymbol{0}}_{R},
\label{eq:initialstatearbitrary}
\end{equation}
in which  the coefficients $c_{01}$ and $c_{02}$ have been defined in Eq. \eqref{eq:c0} and by the basis kets $\left| e_{j(l)}\right\rangle $ we mean that, all of the qubits are prepared in the ground state except the $j$th ($l$th) qubit that is in the excited state.
Accordingly, the quantum state of the entire system+environment at any time can be written as

\begin{equation}
\begin{aligned}
\ket{\psi(t)}&=\Big( C_1(t)\ket{e_j}+C_2(t)\ket{e_l}+C_3(t)\ket{E_{\cancel{j} \cancel{l}}}\Big)e^{-i\omega_{0}t} \ket{\boldsymbol{0}}_{R} \\
 &+\sum_{k}C_{k}(t)e^{-i\omega_k t}\ket{1_{k}}\ket{G},
\end{aligned}
\label{eq:statearbitrary}
\end{equation}
in which we have defined the normalized state $\ket{E_{\cancel{j}\cancel{l}}}:=\frac{1}{\sqrt{n-2}}\sum_{i\neq j,l}^{n}\ket{e_i}$.
Following the procedure presented in obtaining the expression for ${\cal D}(t)$, one may straightforwardly obtain the following analytical expressions for the time-dependent amplitudes
\begin{subequations}
\label{eq:usolved}
\begin{eqnarray}
C_{1}(t)&=&\frac{(n-1)c_{01}-c_{02}}{n}+\frac{c_{01}+c_{02}}{n}{\cal D}(t),  \label{eq:usolved1} \\
C_{2}(t)&=&\frac{(n-1)c_{02}-c_{01}}{n}+\frac{c_{01}+c_{02}}{n}{\cal D}(t), \label{eq:usolved2} \\
C_{3}(t)&=&\frac{\sqrt{n-2}}{n}(c_{01}+c_{02})(-1+{\cal D}(t)), \label{eq:usolved3}
\end{eqnarray}
\end{subequations}
 where ${\cal D}(t)$ has been introduced before. As is stated before, letting $t$  to tend to infinity, then ${\cal D}(\infty)\longrightarrow 0$ which leads to the nonzero values of the coefficients $C_i(\infty)$. Therefore, unlike the case with initial Werner state, the environment not only can create entanglement between various pairs of qubits, but also it may make it to persist to be stationary. According to (\ref{eq:usolved}), this stationary state does not depend on the environment features such as the cavity damping rate or coupling constant as well as the velocity of the qubits but only depends on the initial conditions as well as the size of system, i.e., $n$. This is due to the fact that we have assumed that the coupling constant and the velocity of the qubits be the same for all qubits. It can be shown that, by choosing different coupling constants as well as different velocities associated with different qubits, the stationary entanglement depends also on the cavity damping rate, coupling constants and the velocity as well.

Let us first obtain the expression of the reduced density operator of qubits at any time. This can be done by tracing over the environment variables of (\ref{eq:statearbitrary}) as follows
\begin{equation}
\begin{aligned}
\rho(t)&=\left| C_1(t)\right| ^2\ket{e_j}\bra{e_j}+\left| C_2(t)\right| ^2\ket{e_l}\bra{e_l}+\left| C_3(t)\right| ^2\ket{E_{\cancel{j}\cancel{l}}}\bra{E_{\cancel{j}\cancel{l}}} \\
&+\left( C_1(t)C_2^{*}(t)\ket{e_j}\bra{e_l}+C_1(t)C_3^{*}(t)\ket{e_j}\bra{E_{\cancel{j}\cancel{l}}} \right. \\
&\left. +C_2(t)C_3^{*}(t)\ket{e_l}\bra{E_{\cancel{j}\cancel{l}}} + \text{H.c.} \right)\\
&+\left(1-\left| C_1(t)\right| ^2-\left| C_2(t)\right| ^2-\left| C_3(t)\right| ^2 \right) \ket{G}\bra{G}.
\end{aligned}
 \label{eq:densitymatrix2}
\end{equation}
Based on the above relation, it is possible to consider various pairwise  entanglements resulting from different initial state. For instance, we shall consider the case in which two qubits ($j$th and $l$th) are initially  in a superposition of maximally entangled state (i.e., $s=0$) and when only one qubit (namely $l$th qubit) is initially in the excited state (i.e., $s=+1$).

\subsection{Maximum Entangled Initial State}\label{sec:maxini}
In this subsection, we assume that the system of qubits be initially in a maximum entangled state of two qubits (namely $j$th and $l$th). By tracing of \eqref{eq:densitymatrix2} over all other qubits, we obtain the following reduced density operator
\begin{equation}
\begin{aligned}
\rho_{j,l}(t)&=\left| C_1(t)\right| ^2\ket{e,g}\bra{e,g}+\left| C_2(t)\right| ^2\ket{g,e}\bra{g,e} \\
&+ C_1(t)c_2^{*}(t)\ket{e,g}\bra{g,e}+C_1^{*}(t)C_2(t)\ket{g,e}\bra{e,g} \\
&+\left(1-\left| C_1(t)\right| ^2-\left| C_2(t)\right| ^2 \right) \ket{g,g}\bra{g,g},
\end{aligned}
 \label{eq:rkl}
\end{equation}
which leads to the concurrence
\begin{equation}
{\cal C}_{j,l}(t)=2\left| C_1(t)\right| \left| C_2(t)\right| .
\label{eq:ckl}
\end{equation}
\begin{figure}[h]
\centering
\subfigure[\label{Fig8a} \ Bad cavity limit, $R=0.1$.]{\includegraphics[width=0.23\textwidth]{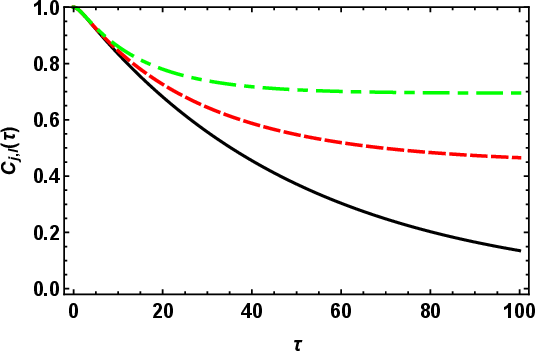}}
\hspace{0.001\textwidth}
\subfigure[\label{Fig8b} \ Good cavity limit, $R=10$.]{\includegraphics[width=0.23\textwidth]{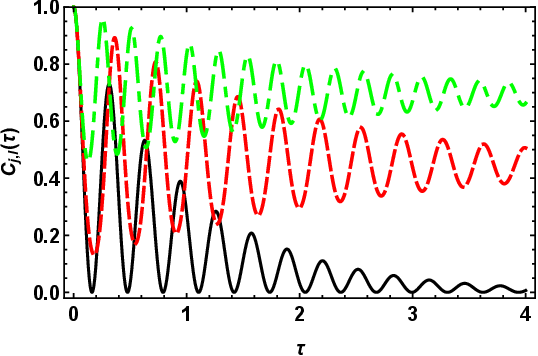}}

\caption{(Color online) Pairwise concurrence ${\cal C}_{j,l}$ as function of $\tau$ for $s=\varphi=0$ with zero velocity in the bad cavity limit, i.e. $R=0.1$ (left plots) and good cavity limit, $R=10$ (right plots) with $n=2$ (solid black lines), $n=6$ (dashed red lines),  and $n=12$ (dot-dashed green lines).} \label{Fig8}
   \end{figure}

Figure \ref{Fig8} illustrates the time evolution of the concurrence ${\cal C}_{j,l}(\tau)$ as a function of the scaled time $\tau$ when the qubits are at rest, i.e., $\beta=0$, for weak and strong coupling regimes for a maximally entangled initial state (i.e., $s=0$ and $\varphi=0$). In the weak coupling regime, concurrence falls down from its maximum initial value and monotonically decreases until it reaches its stationary value. In the strong coupling regime, an oscillatory behaviour along with decaying of entanglement is clearly seen such that for $n=2$ the entanglement sudden death is occurred. As is mentioned before, both strong and weak coupling regimes lead to the same stationary state. The surprising aspect here is that for $n=2$ the entanglement between two qubits vanishes under the environment, but adding more number of qubits maintains the entanglement stored between these two qubits. In general, when the system size $n$ becomes larger, the stationary entanglement increases and the concurrence achieves sooner its stationary value.

\begin{figure}[h!]
\centering
\subfigure[\label{Fig9a} \ Bad cavity limit, $R=0.1$.]{\includegraphics[width=0.23\textwidth]{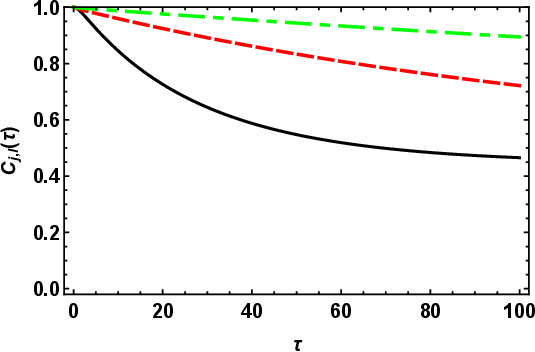}}
\hspace{0.001\textwidth}
\subfigure[\label{Fig9b} \ Good cavity limit, $R=10$.]{\includegraphics[width=0.23\textwidth]{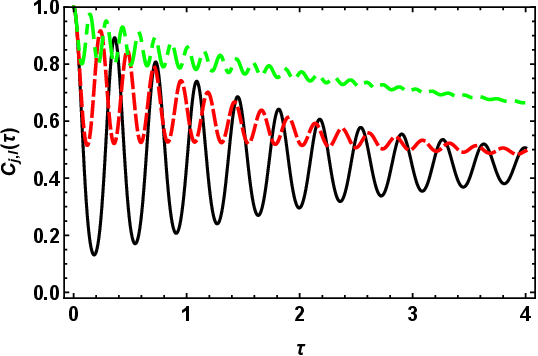}}

\caption{(Color online)  ${\cal C}_{j,l}$ as function of $\tau$ for system size $n=6$ in the bad cavity limit, i.e. $R=0.1$ (left plots) and good cavity limit, $R=10$ (right plots) with $\omega_0/\lambda=1.5\times 10^9$ and ($i$)  $\beta=0$ (solid line), ($ii$)  $\beta=2\times 10^{-9}$ (dashed line), and ($iii$)  $\beta=4\times 10^{-9}$ (dot-dashed line). Other parameters are similar to Fig. \ref{Fig8}.} \label{Fig9}
   \end{figure}

In Fig. \ref{Fig9} we examine the effect of the velocity of qubits on the pairwise entanglement ${\cal C}_{j,l}$  when the  size of the system is $n=6$. Again, the movement of the qubits has a remarkable effect on the survival of the initial entanglement. As is seen, the stationary entanglement does not depend on the velocity of the qubits. Actually, letting $t$ to tend to infinity in Eq. (\ref{eq:ckl}), we found the behaviour of stationary entanglement versus system size $n$ as
\begin{equation}\label{eq:stackl}
{\cal C}_{j,l}(\infty)=\dfrac{(n-2)^2}{n^2}.
\end{equation}
It is clear that, for large values of $n$, ${\cal C}_{j,l}(\infty)$ tends to one.

\begin{figure}[h!]
\centering
\subfigure[\label{Fig10a} \ Bad cavity limit, $R=0.1$.]{\includegraphics[width=0.23\textwidth]{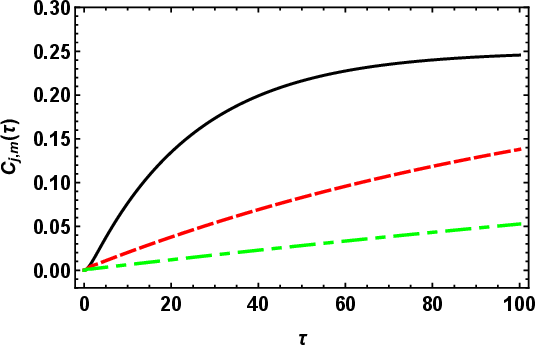}}
\hspace{0.001\textwidth}
\subfigure[\label{Fig10b} \ Good cavity limit, $R=10$.]{\includegraphics[width=0.23\textwidth]{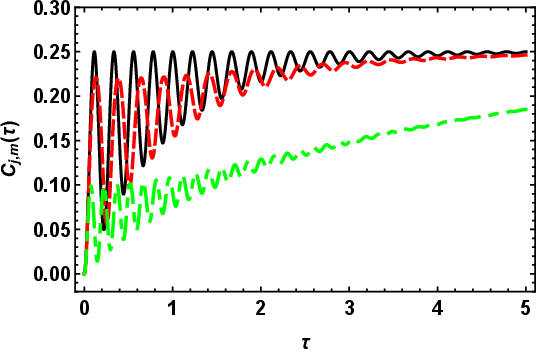}}

\caption{(Color online)  ${\cal C}_{j,m}$ as function of $\tau$ for system size $n=6$ in the bad cavity limit, i.e. $R=0.1$ (left plots) and good cavity limit, $R=10$ (right plots) with $\omega_0/\lambda=1.5\times 10^9$ and ($i$)  $\beta=0$ (solid line), ($ii$)  $\beta=2\times 10^{-9}$ (dashed line), and ($iii$)  $\beta=4\times 10^{-9}$ (dot-dashed line). Other parameters are similar to Fig. \ref{Fig8}.} \label{Fig10}
   \end{figure}

The other possible pairwise entanglement we can study is the entanglement between the $j$th qubit (which is initially in the excited state) and a generic qubit $m$ (which is initially in the ground state). With the help of Eq. (\ref{eq:densitymatrix2}) one can compute the corresponding reduced density operator and obtain the following relevant concurrence:
\begin{equation}
{\cal C}_{j,m}(\tau)=\frac{2}{\sqrt{n-2}}\left| C_1(t)\right| \left| C_3(t)\right|,
\label{eq:ckj}
\end{equation}
which is valid for $n>2$. Figure \ref{Fig10} illustrates the pairwise entanglement  ${\cal C}_{j,m}(\tau)$ for system size $n=6$ for various values of the velocity of qubits. The parameter concurrence starts from its initial value, i.e, zero, as is expected and tends to its stationary value. The movement of the qubits makes the concurrence reaches its stationary value at longer times.  In the good cavity limit and with nonzero values of the velocity of the qubits, the oscillation of the entanglement disappears. Again, the stationary value of the entanglement does not depend on the movement of the qubits.  This stationary entanglement can be determined from Eq. (\ref{eq:ckj}) by letting $t$ going to infinity as follow
\begin{equation}\label{eq:stackj}
{\cal C}_{j,m}(\infty)=\dfrac{2(n-2)}{n^2}.
\end{equation}
It is obvious that, the maximum  stationary entanglement ${\cal C}_{j,m}^{(\text{max})}=0.25$ is achieved for system size $n=4$.

 Finally, the other possible case which we study is the entanglement between two generic qubits $k$ and $m$ initially in the ground state  ($k,m\neq j,l$). The corresponding concurrence reads as
 \begin{equation}
 {\cal C}_{k,m}(t)=\frac{2}{n-2}\left| C_3(t)\right|^2 ,
 \label{eq:cjm}
 \end{equation}
 which is valid for $n>2$.

\begin{figure}[h!]
\centering
\subfigure[\label{Fig11a} \ Bad cavity limit, $R=0.1$.]{\includegraphics[width=0.23\textwidth]{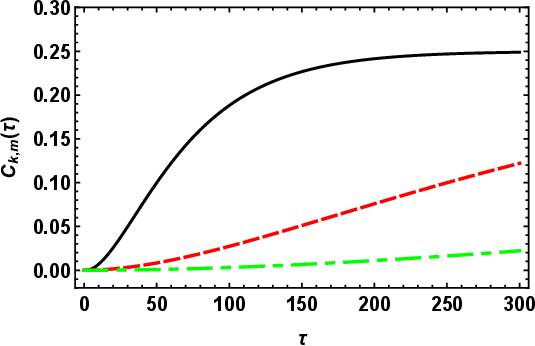}}
\hspace{0.001\textwidth}
\subfigure[\label{Fig11b} \ Good cavity limit, $R=10$.]{\includegraphics[width=0.23\textwidth]{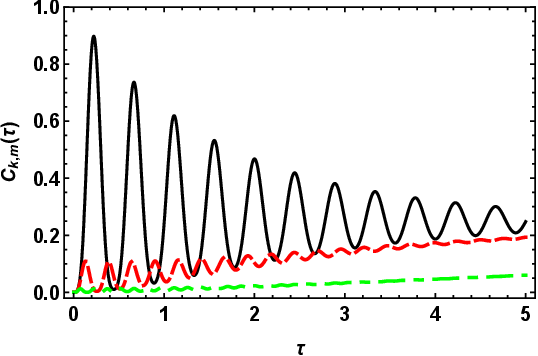}}

\caption{(Color online)  ${\cal C}_{k,m}$ as function of $\tau$ for system size $n=4$ in the bad cavity limit, i.e. $R=0.1$ (left plots) and good cavity limit, $R=10$ (right plots) with $\omega_0/\lambda=1.5\times 10^9$ and ($i$)  $\beta=0$ (solid line), ($ii$)  $\beta=2\times 10^{-9}$ (dashed line), and ($iii$)  $\beta=4\times 10^{-9}$ (dot-dashed line). Other parameters are similar to Fig. \ref{Fig8}.} \label{Fig11}
\end{figure}

 Figure \ref{Fig11} provides the dynamical behaviour of the ${\cal C}_{k,m}(t)$ in the bad and good cavity limits for system size $n=4$ in the presence of the movement of the qubits. It is evident that the entanglement sudden death phenomenon has occurred in the good cavity limit for small values of scaled time $\tau$. It is apparent from the information supplied that, in the latter regime, the amount of revived entanglement has become considerably comparable to 1 at short times. It can be shown that, it is comparable to 1 for small system sizes.

It is also interesting to notice that, both coupling regimes lead to the same stationary entanglement which is independent of the velocity of the qubits. In fact, by letting $t$ to go to infinity in Eq. \eqref{eq:cjm} and computing the stationary concurrence as ${\cal C}_{k,m}(\infty)=\frac{4}{n^2}$, one can easily observe that, for large system sizes, the latter concurrence is by far more negligible compared to the other stationary concurrences.

\subsection{One Initial Excitation}{\label{sec:oneini}
In this subsection, we assume that only $l$th qubit is initially in the excited state (i.e., $s=-1$). In order to analyse the pairwise entanglement between qubit $l$ and another generic qubit $m$, it is enough to set $s=-1$ in  Eqs. (\ref{eq:ckl}) or (\ref{eq:ckj}). The dynamical behaviour of ${\cal C}_{l,m}(\tau)$ is shown in Fig. \ref{Fig12} in both strong and weak coupling regimes for system size $n=4$ for some values of velocity of qubits. It is easy to show that at the steady state, the pairwise concurrence takes the form
\begin{equation}\label{eq:stackjs-1}
{\cal C}_{l,m}(\infty)=\dfrac{2(n-1)}{n^2}.
\end{equation}
On the other hand, the entanglement between two other generic qubits, initially in the ground state, has similar behaviour to Fig. \ref{Fig11}. In particular, its corresponding stationary concurrence takes the form ${\cal C}_{k,m}(\infty)=\frac{2}{n^2}$ which vanishes for large system sizes.

\begin{figure}[h!]
\centering
\subfigure[\label{Fig12a} \ Bad cavity limit, $R=0.1$.]{\includegraphics[width=0.23\textwidth]{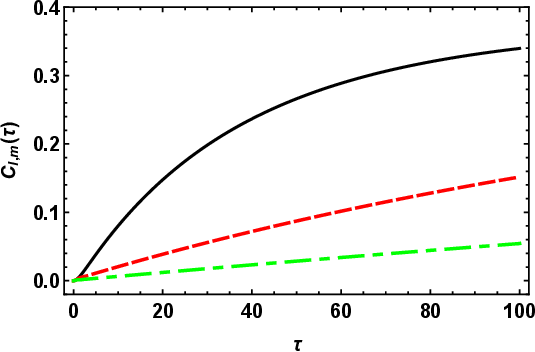}}
\hspace{0.001\textwidth}
\subfigure[\label{Fig12b} \ Good cavity limit, $R=10$.]{\includegraphics[width=0.23\textwidth]{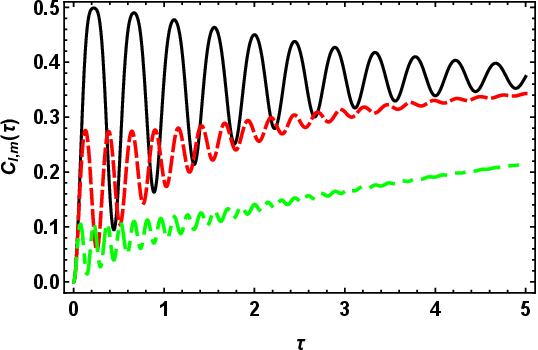}}

\caption{(Color online)  ${\cal C}_{l,m}$ as function of $\tau$ for system size $n=4$ and $s=-1$ in the bad cavity limit, i.e. $R=0.1$ (left plots) and good cavity limit, $R=10$ (right plots) with $\omega_0/\lambda=1.5\times 10^9$ and ($i$)  $\beta=0$ (solid line), ($ii$)  $\beta=2\times 10^{-9}$ (dashed line), and ($iii$)  $\beta=4\times 10^{-9}$ (dot-dashed line).} \label{Fig12}
\end{figure}

Altogether, by comparing various stationary entanglements which have been appeared, it can be concluded that when the system of qubits  initially is in the maximum entangled state of two qubits, we have the graph depicted in Fig. {\ref{Fig13a} as the steady state. The ticker line in this Fig.  implies the fact that, at the steady state, the correlation between the initially excited qubits is stronger than the correlation between any other two qubits. On the other hand, when there is only one excitation in the initial state,  we have a star graph as the steady state (see Fig. \ref{Fig13b}).

\begin{figure}[h!]
\centering
\subfigure[\label{Fig13a} \ Werner initial state.]{\includegraphics[width=0.23\textwidth]{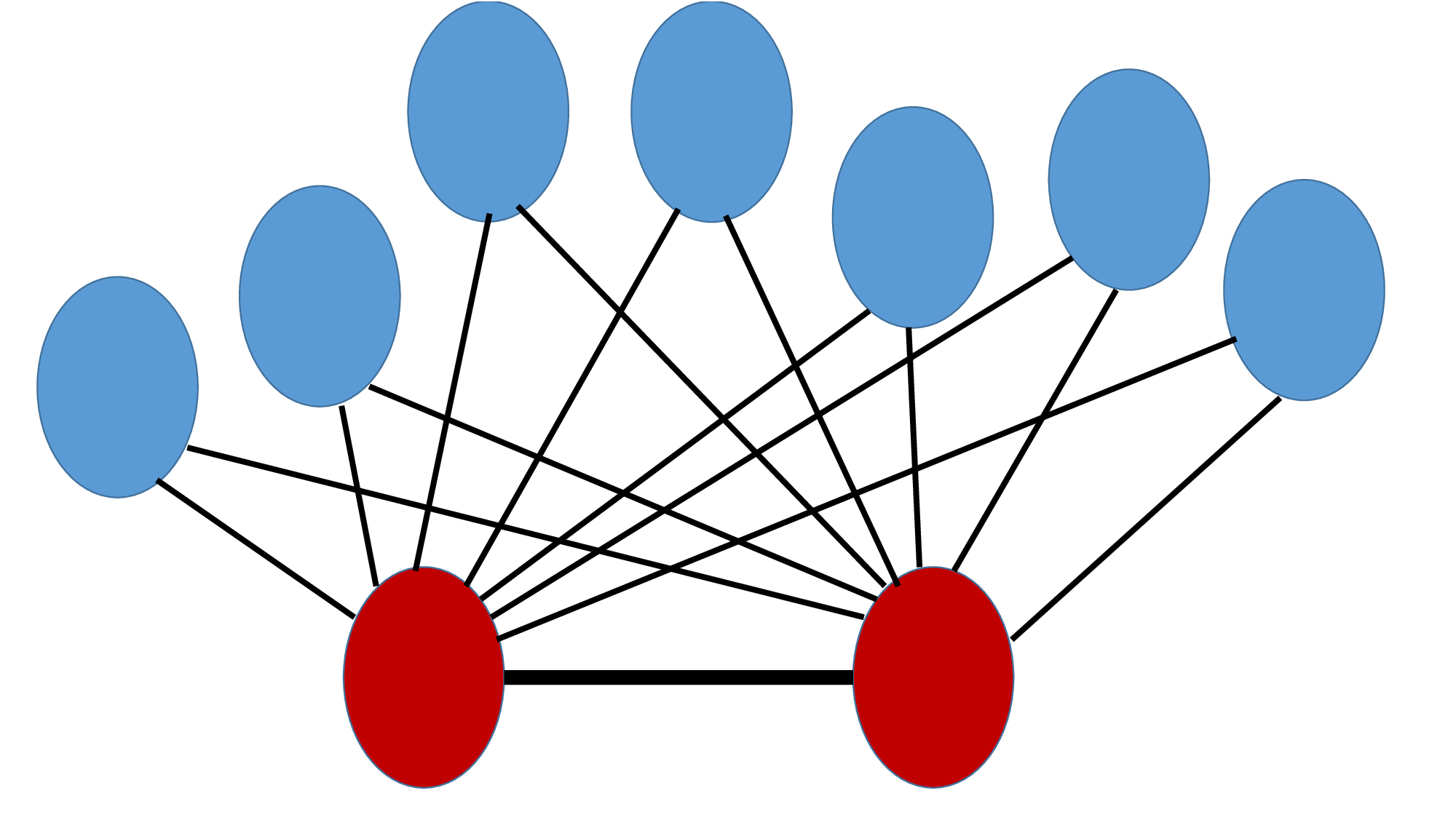}}
\hspace{0.001\textwidth}
\subfigure[\label{Fig13b} \ One excited initial state.]{\includegraphics[width=0.23\textwidth]{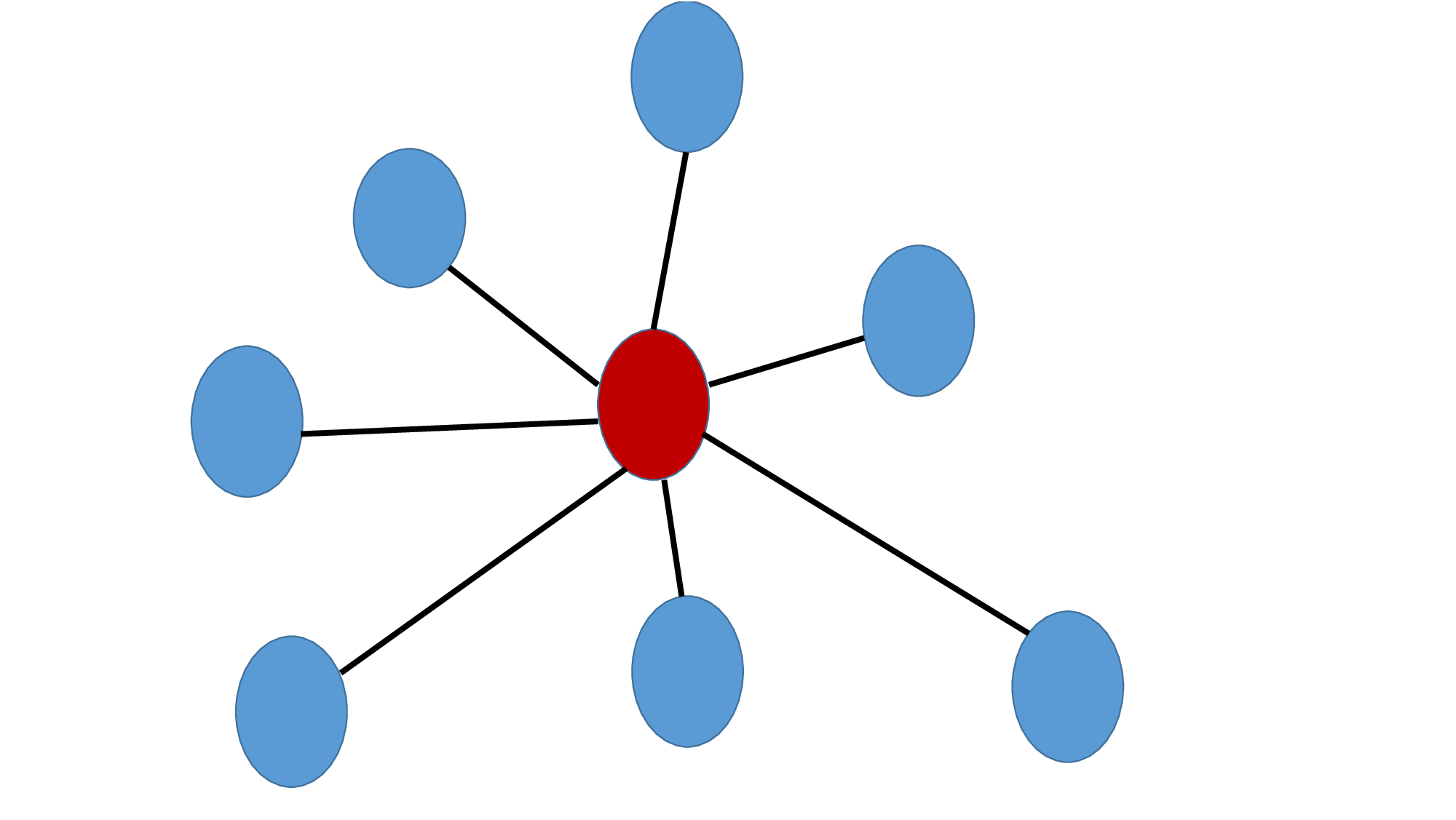}}

\caption{(Color online)   Pictorial representation of the leading stationary concurrence when (a) the initial state is a Werner state and (b) initially only one qubit is in the excited state. Solid lines represent the quantum correlations between qubits at steady state. Red (blue) circles represent qubits initially in the excited states (ground states).} \label{Fig13}
\end{figure}


\section{Conclusion}
\label{sec5}

To sum up, we have considered the problem of entanglement dynamics of moving qubits inside a common environment. The advantage of our model is the consideration of non-Markovian evolution of the moving qubits. The strong coupling of qubits with the environment induces the oscillation of entanglement due to the memory effect of the environment. Whereas, in the weak coupling regime, the pairwise entanglement decays (and sometimes increments) exponentially and goes up to its stationary value asymptotically.

We began with the problem of two moving qubits in a common (global) environment. We have investigated this problem in details. We have found an analytical expression for the concurrence in the presence of qubits movement. We have observed that, there exists a stationary entanglement between these two qubits. The surprising aspect here is that, the stationary state does not depend on the movement of the qubits. This stationary state is exactly the same as state that has been obtained for two qubits at rest \cite{Nourmandipour2015,Maniscalco2008}. We have determined the situations in which a maximally stationary entanglement can be achieved. We then examined the effect of movement on the entanglement dynamics of these two qubits. Our results illustrate that, the movement has a constructive role on the survival of the entanglement in both weak and strong coupling regimes.

We then extended our model into a model consisting of an arbitrary number of moving qubits in a common environment. We did this task in two different ways. First, we considered the case in which the qubits are initially in a Werner state, i.e., a state in which all qubits have the same probability of being in the excited state. We obtained an analytical expression for pairwise concurrence between two arbitrary qubits.  We found that the pairwise concurrence depends directly on the survival probability of the initial state. The pairwise entanglement has a decaying behaviour with no stationary value for both strong and weak coupling regimes. However, by comparing the Figs. \ref{Fig6} and \ref{Fig7} it is evident that, the presence of  movement can preserve the entanglement initially stored in the system of qubits. This is quite similar to the effect of Quantum Zeno effect on the entanglement dynamics of the qubits \cite{Nourmandipour2016}.

Second, since in the case of a Werner state as the initial state of the system of qubits no stationary entanglement can be achieved, we examined the case in which the number of qubits initially in the superposition of only one qubits is limited. This leads to a considerable amount of stationary entanglement. We distinguished between two cases: (i) the system is initially in superposition of one excitation of two arbitrary qubits, and (ii) when only one qubit is initially in the excited state.
In both cases, when the velocity of all qubits is the same, there exists a stationary state which is completely independent of that velocity as well as the environment properties and depends  on the system size $n$ and the initial conditions which is characterized by the separability parameter $s$. Again, the stationary state is similar to the case of non-moving qubits \cite{Nourmandipour2016}.

In the former case, although the interaction of the system-environment leads to vanishing of the entanglement for system size $n=2$ with Bell state as its initial state, as an interesting result, increasing the number of qubits satisfactorily preserves the initial entanglement (see Fig. \ref{Fig8}). Again, the movement of  qubits has a remarkable effect on the survival of the initial entanglement (Fig. \ref{Fig9}). For the strong coupling regime and in the presence of the movement of the qubits, the oscillating behaviour of pairwise entanglement has been vanished. The stationary pairwise entanglement ${\cal C}_{j,l}(\infty)$ (here $j$th and $l$th qubits  are initially in the superposition of one excitation, see Eq. (\ref{eq:statearbitrary})) monotonically increases with the system size $n$ such that for large values of $n$ it tends to 1.

It is also possible to create entanglement for pairs of initially excited and non-excited qubits (see Fig. \ref{Fig10}). As is observed from Fig. \ref{Fig10}, the entanglement can persist at its stationary state which depends only on the system size of qubits. The stationary state is comparable to 1. Again, the movement of  qubits makes the entanglement to reach its stationary value at longer times. It is also possible to create entanglement between pairs of qubits initially in the ground state (see Fig. \ref{Fig11}). However, in such a case, the amount of entanglement is negligibly smaller than the previous cases and also is nearly independent of separability parameter.

In the latter case, when only one qubit is initially in the excited state (i.e., $s=1$ or $-1$), it is possible to generate the entanglement between this qubit and another generic qubit which is initially in the ground state. This amount of this entanglement is comparable to one. Again, the velocity of the qubits makes the entanglement reaches its stationary state at longer times.  In this case, we are left with a star graph as the steady state for large systems in both weak and strong coupling regimes. This is quite in consistent with previous works (see for example, \cite{Nourmandipour2016,Memarzadeh2013}). On the other hand, when two qubits  are initially in a maximum entangled state, we are left with a bipartite graph with strong correlation between the two qubits (which  are initially in a maximum entangled state). On the other hand, previous studies illustrate that when two quits are initially in the excited states simultaneously, at the steady state is a bipartite graph but without any correlation between the two qubits which  are initially excited \cite{Memarzadeh2013}. Altogether, this subject can be of interest from the perspective of quantum complex networks \cite{Perseguers2010}.

The observed aspects in this paper reveal another interesting result, too. By comparing the effects of quantum Zeno on the entanglement dynamics of the qubits in a common environment \cite{Nourmandipour2015,Nourmandipour2016,Nourmandipour2016J,Maniscalco2008}, one can easily observe a similarity between the effect of the velocity of the qubits and the effect of the quantum Zeno on the entanglement dynamics of the qubits.  In \cite{garcia2017entanglement}, the authors have observed such similarity between quantum Zeno effect and the (relativistic) motion of the qubits. This arisen a question that, is there any connection between the velocity of the qubits and quantum Zeno effect? This is left for future works.

Finally, it should be emphasized that our results can be helpful in designing experiments for quantum computation applications when the velocity of the qubits and also the environmental effects cannot be neglected. For instance, our $N$ moving qubits can be modelled as $N$ two-level atoms coupling to the field of 1D photonic waveguide with a  spatially periodic modulation \cite{Calaj2017}. Furthermore, it is possible to implement our model in a circuit QED architecture using a single-mode transmission line resonator interacting with two (or more) superconducting qubits \cite{PhysRevLett.106.030502}.

\appendix

\section{The proof of Eq. (\ref{eq:surv})} \label{App:A}

As is stated before, for the case in which two moving qubits (with the same velocity) are interacting with a common environment, there exists a subradiant state which does not evolve in time (see Eq. \eqref{eq:subradstate}).  The only relevant time evolution is that of its orthogonal, superradiant state
\begin{equation}
\ket{\psi_+}=r_1\ket{e,g}+r_2\ket{g,e}.
\label{eq:superradiantstate}
\end{equation}
The time evolved of the super-radiant state is
\begin{equation}
 \ket{\psi_+(t)}=c_1(t)\ket{e,g}+c_2(t)\ket{g,e}.  \label{eq:superradiantstatetime}
  \end{equation}
In the following, we obtain the survival amplitude ${\cal E}(t)$ of the above state. First, according to \eqref{eq:superradiantstate} and \eqref{eq:superradiantstatetime}, the surviving amplitude is
 \begin{equation}
 \begin{aligned}
{\cal E}(t)&\equiv \braket{\psi_+}{\psi_+(t)} \\
&=\left( r_1\bra{e,g}+r_2\bra{g,e}\right) \left( c_1(t)\ket{e,g}+c_2(t)\ket{g,e}\right) \\
&=r_1c_1(t)+r_2c_2(t)
\end{aligned}
 \label{eq:statesur}
 \end{equation}
By taking derivative with respect to $t$ from above equation, we arrive at
\begin{equation}
\dot{{\cal E}}(t)=r_1\dot{c}_1(t)+r_2\dot{c}_2(t)
\end{equation}
Then using equations (\ref{eq:u1}) and (\ref{eq:u2}), we have
 \begin{equation}
 \begin{aligned}
\dot{{\cal E}}(t)&= -\int_{0}^{t}\! r_1F(t,t')\left( \alpha_1^2c_1(t')+\alpha_1\alpha_2c_2(t')\right) \, \mathrm{d}t' \\
&-\int_{0}^{t}\! r_2F(t,t')\left( \alpha_2^2c_2(t')+\alpha_1\alpha_2c_1(t')\right) \, \mathrm{d}t' ,
\end{aligned}
 \end{equation}
which can be written as
 \begin{equation}
 \begin{aligned}
\dot{{\cal E}}(t)&= -\int_{0}^{t}\! F(t,t')\left( c_1(t')(r_1\alpha_1^2+r_2\alpha_1\alpha_2)\right. \\
&+\left. c_2(t')(r_1\alpha_1\alpha_2+r_2\alpha_2^2\right) \, \mathrm{d}t' ,
\end{aligned}
 \end{equation}
according to the definition of $\alpha_{_T}$ i.e., $\alpha_{_T}=(\alpha_1^2+\alpha_2^2)^{1/2}$ we have
 \begin{equation}
\dot{{\cal E}}(t)= -\alpha_{_T}^2\int_{0}^{t}\! F(t,t')\left( r_1c_1(t') +r_2c_2(t')\right) \, \mathrm{d}t'
 \end{equation}
 Then, using  \eqref{eq:statesur}, we have
  \begin{equation}
 \dot{{\cal E}}(t)= -\alpha_{_T}^2\int_{0}^{t}\! F(t,t')\varepsilon(t^{'})  \, \mathrm{d}t' .
  \end{equation}

 \section*{References}
  \bibliographystyle{apsrev4-1}
  \bibliography{mybib}

\end{document}